\documentstyle[prb,preprint,aps]{revtex}

\begin{document}
\begin{center}

{\Large\bf Electric field and potential around localized scatterers in 
thin metal films studied by scanning tunneling potentiometry}
          
\vspace{0.75cm}
Geetha Ramaswamy\footnote[1]{email: geetha@physics.iisc.ernet.in}\\  
\vspace{0.25cm}
{\small Department of Physics, Indian Institute of Science, Bangalore 560 012, India}\\
\vspace{0.5cm}
A. K. Raychaudhuri\footnote[2]{email: arup@csnpl.ren.nic.in}\\
\vspace{0.25cm}
{\small National Physical Laboratory, New Delhi 110 012, India.}
\end{center}

\begin{abstract}
Direct observation of electric potential and field variation 
near local scatterers like grain boundaries, triple points and 
voids in thin platinum films studied by scanning tunneling 
potentiometry is presented. The field is highest at a void, 
followed by a triple point and a grain boundary. The local 
field near a void can even be four orders of magnitude higher 
than the macroscopic field. This indicates that the void is 
the most likely place for an electromigration induced failure. 
The field build up near a scatterer strongly depends on the 
grain connectivity which is quantified by the average grain 
boundary reflection coefficient, estimated from the resistivity.
\end{abstract}

%\begin{multicols}{2}
\newpage
\tightenlines
Electromigraion (EM) failure is an important reliability 
issue for  metallic interconnects in integrated circuits. 
Increased device density has caused a reduction in 
interconnect size, resulting in very high current densities 
which in turn increase the probability of EM failure. EM 
failures occur either by voids developing across the 
interconnects producing an open circuit or by extrusions
which cause a short circuit with neighbouring lines.
Due to its technological importance, the mechanism of field 
induced void formation has been studied by various techniques 
including microscopy based techniques \cite{ARZT94,PAUL92}. 
Bamboo lines have a nanostructure consisting of a number of 
grain boundaries (GB) spanning the length of the interconnect.
Near bamboo lines contain triple points (TP) and
voids (V) along with a network of GBs. The electric field 
across such a thin film is not uniform at the nanometric scale 
and inhomogeneities in the field and current density occur 
near these localized scatterers \cite{RL57}. The technique 
of scanning tunneling potentiometry (STP) invented by Muralt 
and Pohl \cite{MUR86}, which is built around a scanning 
tunneling microscope (STM), offers a novel method of 
studying the potential variations and hence the electric field 
at these scattering centers with a nanometric resolution, by 
providing a simultaneous map of the topograpy and the potential 
distribution in a current carrying film. Thus it provides 
information regarding the nanostructure and the field 
distribution, both of which are of vital importance in 
EM studies.

In this letter we present our results on the study of 
electronic transport in polycrystalline thin platinum films 
by STM/STP. By simultaneously mapping out the topography and 
the potential distribution in the current carrying thin film, 
we investigate the variation in the potential and the build up 
of an electric field near various types of localized scatterers, 
in the scale of the spatial extent of the scatterers. We show 
that the spatial distribution and the magnitude of the local 
field depends on the type of the scatterer as well as the 
connectivity of the grains. Specifically, our experiments show 
that of the different types of scatterers present in a 
polycrystalline thin film, the maximum build up of an electric 
field occurs near a void, which then is the most likely spot
for an EM induced failure. Further our study also shows that
for a given type of scatterer, the magnitude of the local 
field depends on the grain connectivity and it is low for well
connected grains. 

Previous STP studies on metallic films have focused on 
charging effects in granular gold films \cite{MUR86} and 
on local variations in potential in Au-Pd \cite{KIR88}, 
Au \cite{SNE96} and Bi thin films \cite{FEN196}. However, 
local variations in electric field around various kinds of 
extended defects occurring in a thin film has, so far, not 
been studied. Ideally, the STP experiments should be performed 
on conventional interconnect materials like Al and Cu. 
However, the formation of surface oxides in these materials 
on exposure to ambient atmosphere makes the interpretation of 
STP results difficult. In an attempt to understand the 
microscopic origin of the field inhomogeneities and the 
relative importance of the different kinds of defects in 
inducing an EM failure, we decided to study the underlying 
physical process in platinum, which is resistant to oxidation. 

Platinum films (thickness$\approx$10~nm, deposition rate 
$\approx$2.8~nm/min) were deposited using shadow masks on 
cleaned glass substrates by e-beam in a turbo pumped chamber 
at a base pressure of 5~X~10$^{-8}$ torr, using material of 
6N purity. Simultaneous STM/STP images were obtained by the 
double feedback technique, imaging the films under ambient 
conditions in an STM built in-house, using Pt-Rh(13\%) tips. 
The details of the STM and the double feedback operation have 
been described elsewhere \cite{STM97,THESIS}. The topographic 
images were obtained with a tunnel current of $I_{pp}$~=~0.8~nA 
and an AC bias of $V_{pp}$~=~0.05~V at a frequency of 2~kHz. 
The potentiometic images were obtained with a macroscopic 
field of 5.2~V/cm and a current density {\bf j} $\approx$ 
$10^5$~A/cm$^2$. To avoid any artifacts that might arise if 
the current direction and the scan direction are the same, 
{\bf j} was kept at an angle ($-70^o$) to the fast scan 
direction (X axis).

Figure~1 (a) and (b) show the simultaneous STM and STP images 
respectively, in a 62~nm~X~44~nm region of the film. The 
topographic image shows a number of grains, GB, TP and V. 
Representative ones are marked in the figure. The nanostructure 
consists mostly of circular grains having an average grain 
diameter $<D>$ $\approx$ 14.7~nm and an r.m.s surface roughness 
$\approx$ 1.6~nm, as obtained from several topographic scans 
performed in different regions of the film. Fig.~1(c) is a 
line profile across the topographic image (marked across the 
image) showing the z-height corrugations of the GBs and TPs. 
From the potentiometric image and the line profile shown in 
Fig.~1(d), it is immediately apparent that the potential does 
not drop uniformly across the film surface and that the 
local potential distribution is severely affected in the 
vicinity of the scatterers. The line profile in Fig.~1(d) 
is obtained across the potential image in the same region 
as the profile in Fig.~1(c). From the two line profiles we 
discern a one to one correspondence between the scattering 
centers in the topographic image and the voltage variations 
in the potential image. 

We analysed the STP images obtained from various regions of 
the film to ascertain the magnitude of the potential 
inhomogeneities caused by different types of extended defects 
and their relative importance in contributing to a possible
EM failure. Our analysis shows that the typical potential variations 
$\Delta$$\phi$ at the GB are $\approx$ 1~-~3~mV occurring over a 
range of 0.5~-~3.5~nm. At the TP and voids $\Delta$$\phi$ is 
$\approx$ 1~-~4~mV and 3~-~9~mV respectively, occurring over 
a distance of 1~-~3~nm. It is to be noted that the potential 
image is flat and featureless in the absence of a DC current 
through the sample, except for a few 0.3~mV quantization 
noise wiggles of the A/D converter, as seen from Fig.~1(e). 
This proves that the potential variations are the result of 
an actual build up of a field near the scatterer and are 
not due to tip related artifacts.

In order to further understand the nature of the grain 
connectivity and the extent and magnitude of the scattering 
at different scattering centers such as GB, TP and V, we 
investigated the electric field values in their vicinity. 
The local transport field in the surface plane is calculated 
from the gradient of the local potential, $\phi(x,y)$, 
as:~${\bf{E}_{\parallel}}(x,y)~=~-~{\bf{\nabla}}\phi(x,y)$. 
$\phi(x,y)$ is what is measured in an STP scan. We computed 
the electric fields along the X, Y directions numerically from 
the 128~X~128 potentiometric data array and used quiver plots 
of the field data to visually show the distribution of field 
lines near the scatterer. The results of the field calculation
are shown in Fig.~2. The length of the arrow indicates 
the magnitude and the head points along the direction of the 
field.

Figure.~2 shows the electric field and a line scan across the 
potential image around a GB, a TP and a V. These regions 
have been labeled and marked by a rectangle in Fig.~1(a),
with a short line across the rectangle depicting the profile. 
A comparison of the figures brings out the distinct nature of 
the electric field distribution around the three kinds of 
scatterers. The field lines in a GB (Fig.~2(a)) are 
concentrated along the periphery of the grain and very low
in its interior. At a TP the field lines are stronger 
(Fig.~2(c)). The field radiates outward from the GB and 
concentrates at the TP. At a V, the build up of the field is
most prominent (Fig.~2(e)). This is also brought out by the 
STP line profiles (Figs.~2(b),(d),(f)), which show that 
the potential inhomogeneity is a maximum at a V and minimum 
at a GB. Typical field values obtained from above are 
$\approx$ 1.6~X~$10^4$~V/cm, 3.2~X~$10^4$~V/cm and 
7.2~X~$10^4$~V/cm respectively for a GB, TP and V.  

Our study also indicates that the field build up around 
a scatterer will depend on the grain connectivity for 
a given kind of scatterer. The grain connectivity of the 
film can be characterized by an average GB reflection 
coefficient <R$_{g}$>, which is obtained from an analysis 
of the temperature dependence of resistivity of the film 
\cite{SAM,GR}. From the analysis of the resistivity 
data ($\rho_{4.2K}$$\approx$47$\mu\Omega$cm and 
$\rho_{300K}$/$\rho_{4.2K}$=1.05), we obtain an 
<R$_{g}$>$\approx$0.9 for this film. In order to better
understand the dependence of <$R_{g}$> on the grain connectivity, 
which is reflected in the field build up, we repeated the 
STM/STP experiments on a platinum film which was grown on 
a rougher surface but with similar thickness and average 
nanostructural parameters ($<D>$$\approx$12.35~nm and an 
r.m.s roughness of 1.7nm). The film had a 
$\rho_{4.2K}$$\approx$160$\mu\Omega$cm 
and $\rho_{300K}$/$\rho_{4.2K}$=1.22, resulting in a 
<$R_{g}$>$\approx$0.97. This shows that the grains are not 
as well connected as in the previous film. In this film, 
although the STP and the field patterns are qualitatively similar 
to the previous film, for the same macroscopic field, the 
magnitude of the field at the defects (GB, TP and V) 
are an order of magnitude larger. This clearly indicates 
that the field build up, for a particular kind of defect,
strongly depends on the grain connectivity. 

This study has brought out a number of important observations. 
The quiver plots conclusively prove that the field across the 
film is not uniform and it is severely altered at the 
scatterers. Further, the field tends to concentrate at 
the scatterer and it is very low in the interior of the grain. 
The magnitude of the field, for a given scatterer, depends
on the grain connectivity. Among the different kinds of 
scatterers, the field is highest at a void. The mapping out 
of the potential and field at the different scattering
sites is an important result as this has significant 
implications on our understanding of EM phenomena in an 
interconnect. It identifies regions of excessive field build up, 
which are the likely "weak spots" where an EM failure can 
originate. Previous studies on thin films have shown that
during EM, vacancies migrate predominantly along GBs and 
accumulate at GBTPs due to vacancy flux divergence 
\cite{ROSE,HO}. These vacancies coalesce into a void, 
which then becomes the most likely spot for an EM failure 
to occur, when it reaches a critical size. Our STP 
study supports this observation. If the grain connectivity 
is poor, the local field at a triple point may 
even be three orders of magnitude higher than the average 
macroscopic value. Such a high field can result in a void 
formation, which has a higher field build up and hence 
has the greatest probability for EM induced film failure.

In conclusion, we have established a procedure for identifying 
the hot spots for a field induced EM failure. Our study shows 
that wide line metal interconnects should have well connected 
grains with a minimum number of triple points and voids for 
reliable long term performance. 

We wish to thank Prof. N. Chandrasekhar for allowing us the 
use of his e-beam facility and K. Das Gupta for depositing the 
films.

%\end{multicols}

\newpage
{\centerline {FIGURE CAPTIONS}}            

(1) FIG.1. Simultaneous STM/STP scans in a 62~nm~X~44~nm area. 
(a) topographic scan (b) potentiometric scan. The arrow 
indicates the direction of the macroscopic {\bf j}. 
The profiles across (a) and (b) are shown in (c) and (d)
respectively. (e) STP profile in the absence of a field.

(2) FIG.~2. The field distributions and line profiles
at various types of scatterers, calculated from the STP image, 
obtained in the regions marked by rectangles and short lines 
respectively in Fig.~1(a). (a) and (b) Field and profile 
at a GB, (c) and (d) at a TP and (e) and (f) at a V.
It is seen that the field is maximum at a void, followed by
a TP and a GB and it is very low in the interior of the grain.

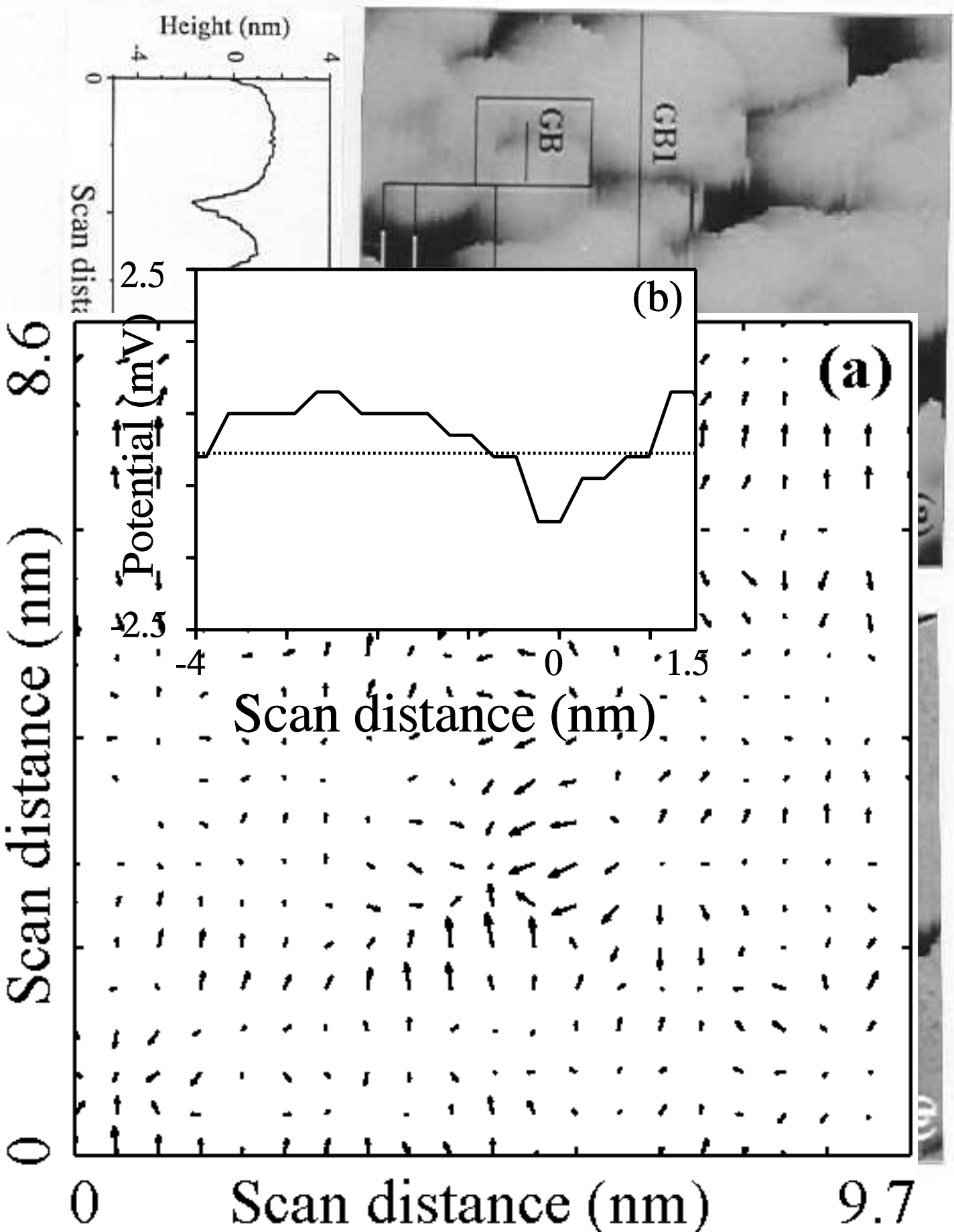
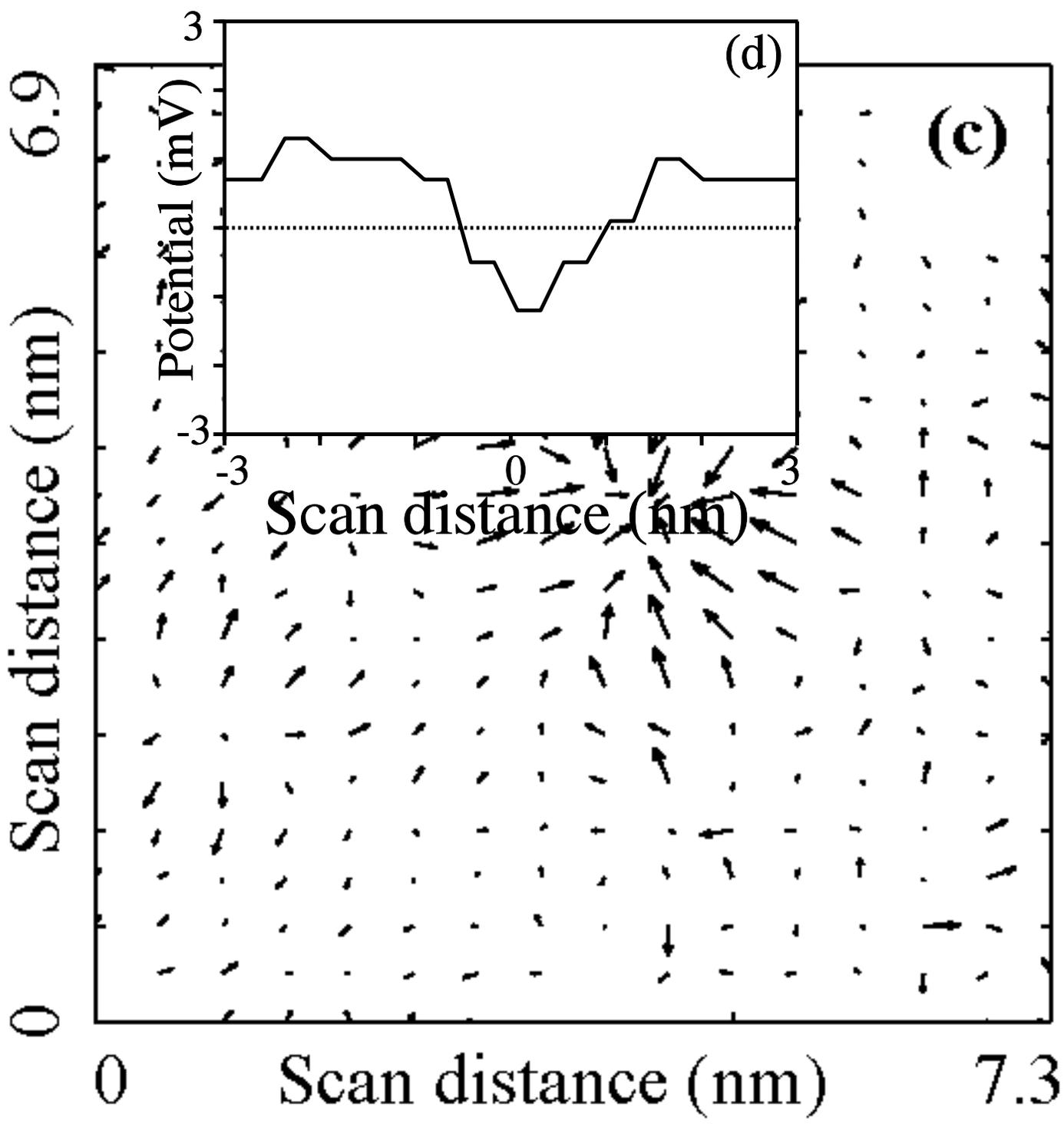
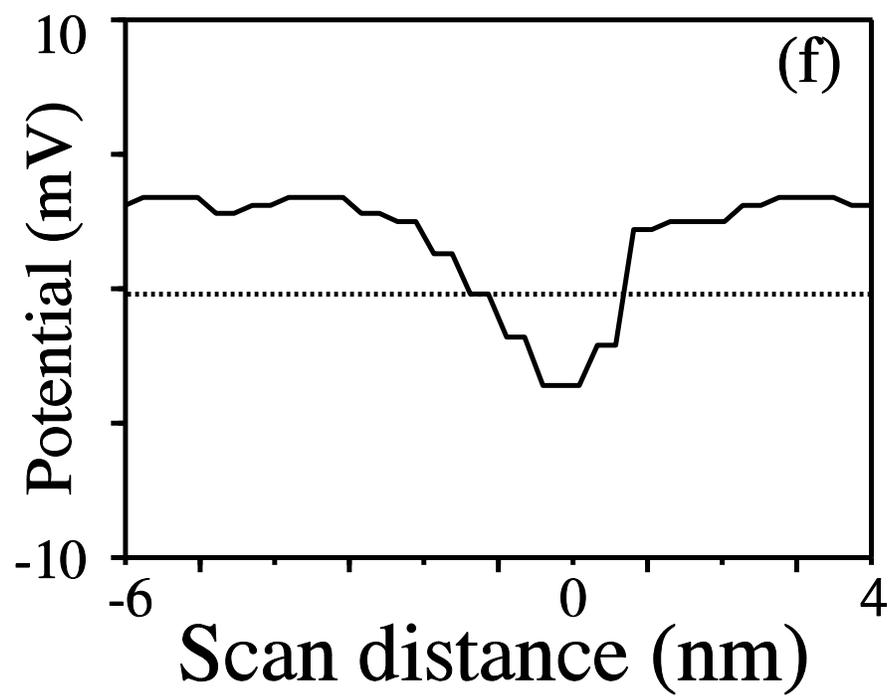
\end{document}